\documentclass[12pt]{article}
\usepackage{epsf}
\def\be{\begin{equation}}
\def\ee{\end{equation}}
\def\beq{\begin{eqnarray}}
\def\eeq{\end{eqnarray}}
\begin{document}
\title{ PICO-CANONICAL ENSEMBLES: A THEORETICAL DESCRIPTION OF
METASTABLE STATES}

\maketitle

\begin{center}
Deepak Dhar\\ 

Department of Theoretical Physics,\\
Tata Institute of Fundamental Research, \\
Homi Bhabha Road, Mumbai  400005 (India)
\end{center}

\abstract{ We define restricted ensembles, called pico-canonical
ensembles, for a statistical-mechanical description of the metastable and
glassy phases. In this approach, time-evolution is Markovian, with
temperature dependent rates. Below a particular glass-temperature, the
system is strongly non-ergodic, and the phase space breaks up
into a large number of mutually disconnected sectors. Averages are
calculated over states within one such sector, and then averaged over
sectors. As a soluble example, we calculate these explicitly for a one
dimensional lattice gas with nearest neighbor couplings.}

\newpage

\section{Introduction}

It is perhaps not an exaggeration to say that the general principles of
determining equilibrium properties of systems are quite well understood by
now. Given a system of $N$ molecules, interacting with each other with a
given pair potential, one calculates the partition function. Various
thermodynamic quantities such as pressure, specific heat can be obtained
by differentiating the free energy with respect to appropriate variables.
When the exact calculation of partition function is not possible,
different approximation schemes such as high- or low- temperature
expansions, mean-field or other variational approximations can be used to
get the qualitative and quantitative behavior.

This procedure, however, fails completely  for metastable phases,
such as diamonds or ordinary glass. This is because, the partition
function, if correctly calculated, would only give properties of
crystalline states of graphite, or quartz. In the standard Bolzmann-Gibbs
prescription, if the system could easily explore all parts of the phase
space available to it, the probability that the system will be found in a
configuration corresponding to the diamond- or the glassy structure, is
negligible. The reason for the failure of the Boltzmann-Gibbs prescription
is non-ergodicity.  The system in the glassy state is said to be trapped
in a local minimum of the free energy \cite{footnote}. Transitions away
from a local neighborhood of these states occur at very slow rate (the
time scale in case of window -glass is centuries).  Once the system gets
out of these trapping states, it is very unlikely to return to them, and
these states do not have a significant weight in the equilibrium state,
which
may be defined as the very-long-time steady state. However, these states
determine the averages of macroscopic quantities measured at the
laboratory time scales, quite different from the ``truly long-time
averages'', calculated correctly by equilibrium statistical mechanics.

Usually, one treats glasses as systems relaxing to equilibrium very
slowly, and there is a lot of interest in the study of such slow
relaxations.  On the other hand, the properties of a piece of diamond, or
glass do not change appreciably over a period of days, or years. An
experimentalist can put them in a calorimeter, and measure the specific
heat, or apply pressure to determine the bulk modulus. But for a theorist,
these are non-equilibrium states of matter, and notions like free energy
are not `well-defined'. In this paper, we show how the conventional
framework of equilibrium statistical mechanics can be extended to include
a description of such systems, and look upon glasses and diamonds, not as
{\it evolving to equilibrium, but as in equilibrium}.

   The macroscopic description of the glassy state should involve only a
small number of variables, and certainly not the precise specification of
all positions and momenta of all the atoms.  Hence we necessarily deal
with only a probabilistic description giving the probability of occurence
of different microstates. However, the ensemble of  available
microscopic states must be restricted to those which are `nearby', and are
accessible to the system within times comparable to the experimental
time-scales. Such an ensemble in these systems is much smaller than the
corresponding microcanonical ensemble. The ratio of number of states in
one such ensemble to that of the standard {\it micro-}canonical ensemble
decreases exponentially with the size of the system. We shall call such
ensembles {\it pico-canonical ensembles}.

   While the general idea outlined above is well-known, and accepted, it
has not been possible so far to define precisely this concept of
restricted ensembles in a way which allows calculation of such partition
functions, and thus determine the averages in any nontrivial case, {\it
with short-ranged interactions}. This is what we shall try to do here for
the simple toy model of Ising spins on a line, with nearest neighbor
interactions, and a non-ergodic dynamics \cite{seoul,gimenon}.  In this
paper, we will argue that this analyically tractable model provides a
useful paradigm for the statistical description of the glassy phase. We
shall not attempt to review here the very large body of work which already
exists dealing with the phenomena of ageing, i.e.  long-time relaxation in
glasses and spin-glasses \cite{others}.

   The key observation here is to note that the processes operating in a
glass may be divided into two classes:  fast and slow. Fast processes are
those whose time scales are much shorter than that of observation, e.g.
the vibratory motion of atoms about their mean positions. Slow processes
are those involving rearrangement of atomic configurations, particle- or
vacancy diffusion, creep etc. To define a quasi-equilibrium state of
glassy systems, we have to imagine that ``all the fast things have
happened'', and the slow things have not even started \cite{feynman}. To
be precise, we postulate that the system dynamics is such that if the slow
processes {\it do not occur at all,} the dynamics becomes {\it nonergodic}
and the phase space breaks up into a large number of {\it disconnected}
pieces, called sectors. The number of such sectors increases exponentially
with the size of the system. We shall call such system many-sector
decomposable (MSD), and in pico-canonical ensembles we sum over only one
of these disconnected sectors.

   We can explicitly compute the partition functions corresponding to such
pico-canonical ensembles in our simple model. These can, and do, vary from
sector to sector. Which one corresponds to the experimental system?  A
precise specification of the sector is not possible in a macroscopic
description. As the number of sectors grows exponentially with system
size, one would need order $N$ binary bits just to characterize the
sector, where $N$ is the number of atoms in the system. In an experiment,
one can, at best, hope to give some specification of how the system
is prepared. In a theoretical calculation,
this implies that in addition to calculating the free energies in
different sectors, given the cooling schedule etc., one has to also
determine the probability distribution that one ends up in one of the many
sectors. Thus, the free energy must then be further averaged over sectors
with a suitable weight for each sector, which will depend on the history
of the system etc.

This averaging is similar to that over quenched disorder in the usual
approach to glasses, but {\it this disorder is a self-generated}. Also, as
we shall see in our model, the ``frozen degrees of freedom'' need not be
frozen in real space.  The relative weight of different sectors is not an
additional input to calculation, but is determined by the model.

\medskip

\section{The Model}

     The model we shall choose to address these issues is a very simple
one. We consider $N$ hard-core point particles on a linear chain of $L$
sites. There is an attractive interaction amongst the nearest neighbor
particles of strength J.  The Hamiltonian of the model is given to be

\beq
      H = -J \sum_{i=1}^L n_i n_{i+1}
\eeq
where $n_i$ is $1$ or $0$ depending on whether the site i is occupied or
unoccupied. Clearly $\sum_i n_i =N$.  We assume that the system evolves in
continuous time  by a local Markovian dynamics. We shall choose the
transition rates to satisfy the detailed balance condition. Then, in the
steady state all accessible configurations occur with their Boltzmann
weights.

     Now for a more detailed specification of the allowed transitions: We
assume that the particles can diffuse to nearby sites either by simple or
by assisted diffusion. In simple diffusion, a particle at site $i$ can
jump to an empty neighbor with a rate $\Gamma_1 exp(- \frac{\Delta
E}{2kT})$, where $\Delta E$ is the change in the energy of the
configuration, and $T$ is the temperature of the system. This may be
represented by the equation

\be
 01 \rightarrow 10 
\ee

In the case of assisted diffusion, two adjacent occupied particles (a
dimer) can jump together one step left or right with a rate $\Gamma_2
exp(- \frac{\Delta E}{2kT})$. Then this process can be represented by a
`chemical' equation

\be
 110  {\leftarrow\rightarrow} 011
\ee

However, these pairings are not transient, and the dimers can
`reconstitute'. Thus, for example, in the sequence of transitions

\be
  ..11010.. {\rightarrow} ..01110.. {\rightarrow} ..01011..
\ee
the middle particle is first paired with particle on the left, and then in 
the second transition with the particle on the right.

\begin{figure}[htb]
\centerline{
        \epsfxsize=11.0cm
        \epsfysize=6.0cm
        \epsfbox{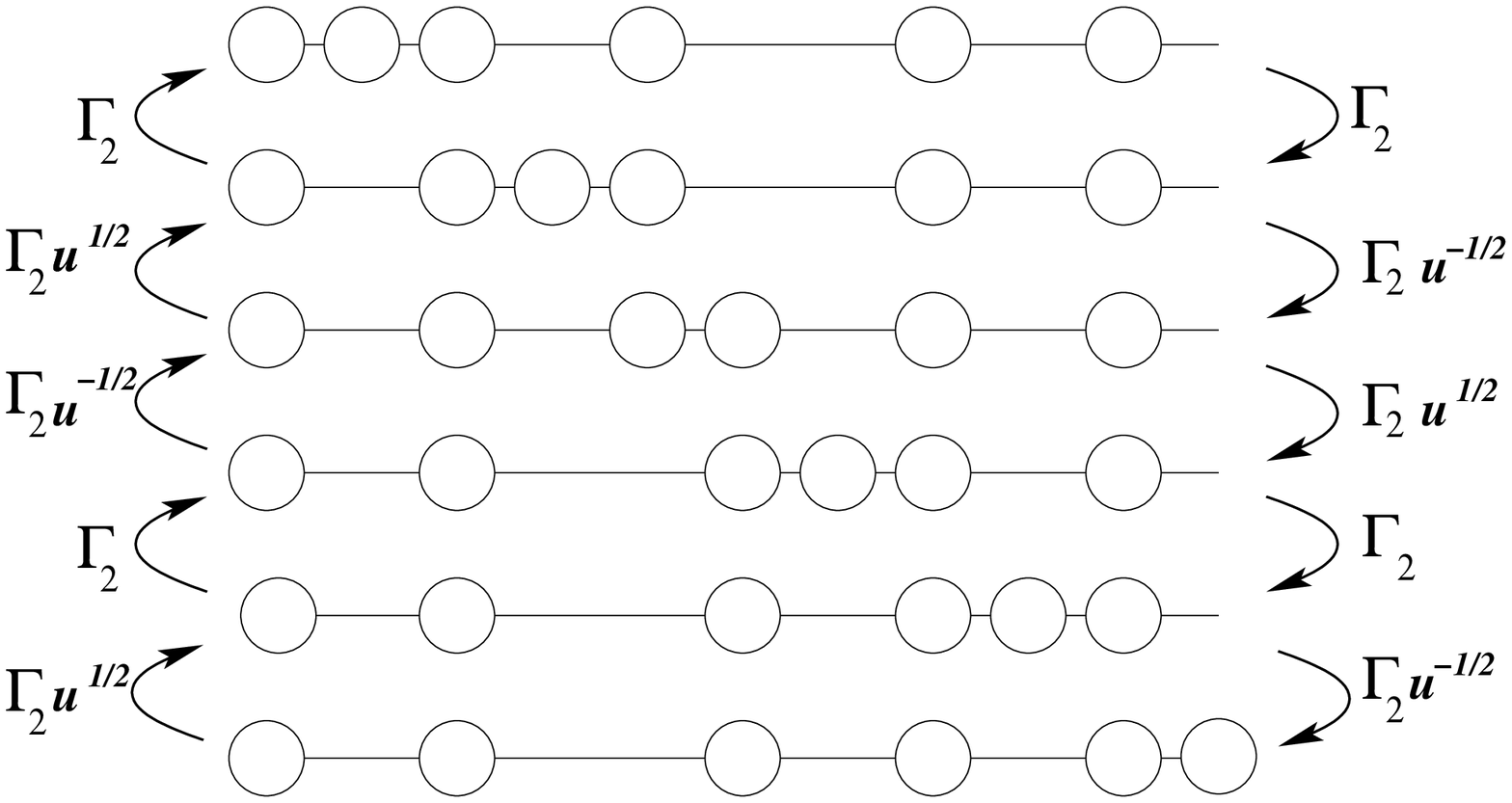}}
\label{fig:1}
\end{figure}

{\bf Fig. 1:} An example of breakup of phase space into sectors when
monomer diffusion is not allowed. There are only six configurations in
this sector of the phase space for $L= 11, N=6$. All configurations have
the same irreducible string $101001010$. The rates of transitions amongst
them are shown. In the steady state, their weights in the partition sum
are  $u^2: u^2: u: u^2: u^2: u$.

Our reaction rates have been chosen to satisfy the detailed balance
condition, and so the steady-state properties of the system do not depend
on the precise values of $\Gamma_1$ and $\Gamma_2$. The probability of
different configurations in the long-time state starting with a given
initial configuration is immediately written down. It is proportional to
$exp(-\beta H)$ for all configurations within the sector in which the
initial configuration lies, and $0$ for others.  We assume that $\Gamma_1$
has a strong dependence on temperature, so that it is effectively zero for
all temperatures $T < T_{G}$, but is non-zero for temperatures greater
that $T_{G}$. $\Gamma_2$ may be assumed temperature -independent. Thus, at
low temperatures, this model reduces to a system of Diffusing
Reconstituting Dimers (DRD) \cite{seoul}

    The assumption that monomers have a much lower diffusivity than dimers
at low temperatures makes the model mathematically tractable, but it is
not unrealistic: for example, platinum dimers on some surfaces have higher
mobility than monomers.

\section{Calculation of the Number and Sizes of Sectors}

For a linear chain of L sites, the phase space consists of $2^L$ distinct
configurations.  The conservation of particle number implies that the
total phase space can be broken into $(L+1)$ disconnected parts, each
corresponding to a different value of the total particle number. For $T >
T_{0}$, it is easy to see that all configurations having the same number
of particles can be reached from any one of them.

However, for $T < T_{0}$, the number of disconnected sectors becomes much
bigger. The number of sectors increases as $exp(L)$. This is easy to see.  
All configurations having only isolated $1$'s in a background of $0$'s,
with no dimers cannot evolve at all. For example,
$\ldots0010100010001\ldots$ cannot evolve, and consitutes a sector, with
only one configuration in the sector. The number of such configurations
increases as $exp(L)$, and provides a lower bound on the total number of
sectors.

The decomposition of phase space into disjoint sectors is described most
simply in terms of a construct called the {\it irreducible string}
\cite{pramana,prl,harimenon}. With each of the $2^L$ possible
configurations of the DRD model, we attach a binary string, called the IS
corresponding to that configuration, constructed as follows: We read the
$L$-bit binary string of ${n_i}$ specifying the configuration from left to
right until the first pair of adjacent $1$'s is encountered. This pair is
deleted, reducing the length of the string by $2$. This process is
repeated until no further deletions are possible. The resulting string is
the IS for the configuration. For example, for the binary string
$01001110110$, the irreducible string is $0100100$.

By construction, for each configuration, there is a unique IS. If a dimer
diffuses, the IS is not changed. Thus the IS is a constant of motion.  It
is easy to show that two different configurations belong to the same
sector, if and only if they both have the same IS. The IS thus provides a
unique label for each sector.

     The DRD dynamics can be viewed as the exclusion process of a system
of 3 species $A$, $B$ and $C$ of particles on a line. In this process each
site of linear chain is occupied by a single particle, which may be of
type $A$, $B$, or $C$. We set up a one-to-one correspondence between the
configurations of the DRD model, and of the exclusion process as follows:
Read the binary string of the DRD from  left to right, and use
the
substitution rules $11 \rightarrow A, 10 \rightarrow B, 0 \rightarrow C$.
The stochastic evolution rule in terms of the exclusion process
configurations is simply that a particle of type $A$ can exchange position
with a particle of type $B$ or $C$ at an adjacent site. Type $B$ and $C$
cannot exchange positions with each other.  This model thus is a special
case of the k-species exclusion process \cite{boldri,derrida,ferrari}.

   The conservation of IS in this language corresponds to the simple
statement that as $B$ and $C$ type particles cannot exchange places, their
relative order is unchanged in time. Starting with an initial
configuration specified by a string composed of three characters $A$, $B$
and $C$, deleting all the occurrences of $A$'s in the string, we are left
only with a string composed of $B$'s and $C$'s which is conserved by the
dynamics.
    
    In terms of the exclusion process, it is quite straightforward to
write down the formulas for the number of distinct sectors for the DRD
model, and also the number of configurations in each sector. In a sector
having $N_A$ particles of type $A$, the length of the IS is $L- 2 N_A$.
The number of unpaired $1$'s is $N_B = N - 2 N_A$, and the number of type
C particles is $N_C = L - 2 N + 2 N_A$.

 The number of configuration within a sector with a specified IS, say
${\cal I} = B C C C B B C B C \ldots$, is the number of ways we can place
$N_A$ $A$'s between $(N_B + N_C)$ $B$'s and $C$'s. So this number is

\begin{eqnarray}
\nonumber
    \Omega (N_A~|~{\cal I}) &=& (N_A + N_B+ N_C)!/[ N_A! (N_B + N_C)!]\\
                         &=&(L - N +N_A)!/[N_A! ( L -N)!]  
\end{eqnarray}

The number of different sectors having a given value of $N_A$ is the
number of distinct ways of writing the IS consisting of $N_B$ $B$'s and
$N_C$ $C$'s.  This number is $(N_B + N_C)!/(N_B! N_C!)$. The numbers $N_B$
and $N_C$ are known once $N_A$ is known.  To calculate the total number of
sectors, we have to sum the number of sectors  for 
$0 \leq N_A \leq N/2$.

\section{Calculating Partition Functions}

Let ${\cal Z}( N, L ~| ~ {\cal I})$ be the partition function for a linear
chain with $L$ sites with $N$ atoms, in a sector corresponding to IS
${\cal I}$. Then ${\cal Z}( N, L ~| ~ {\cal I})$ is a polynomial in $u =
exp(\beta J)$.  We define the generating function
\be
{\tilde {\cal Z}}(x,z,u~|~{\cal I})= \sum_{L=1}^{\infty}\sum_{N=0}^{L}
x^L z^N {\cal Z}(L,N~|~{\cal I})
\ee

It is easy to see that the generating function for the IS ${\cal I}'= X
{\cal I}$, where $X = B$ or $C$ factorizes simply as
\be
{\tilde {\cal Z}}(x,z,u~|~ X{\cal I}) = w_X {\tilde {\cal Z}}(x,z,u~|~ {\cal I})
\ee

where $w_B$ sums over substrings reducible to $B$, $i.e.$ $10 + 1110 +
1111110 \ldots$, giving 
\be 
w_B = x^2 z u /( 1 - x^2 z^2 u^2) 
\ee

Similarly, we have \be w_C = x + x^3 z^2 u /( 1 - x^2 z^2 u^2) \ee If the
number of $B$'s and $C$'s in ${\cal I}$ is $N_B$ and $N_C$ respectively,
we have \be {\tilde {\cal Z}}(x,z,u~|~ {\cal I}) = w_B^{N_B } w_C^{ N_C
+1}/x \label{Ezis} \ee As a simple check, we see that the generating
function for the null string is $w_C/x$.

For temperatures above $T_G$, we have to sum over all possible IS. This is
easy. We note that a formal series over all possible string of $B$'s and
$C$'s can be formally summed as

\be
\sum_{n=0}^{\infty} ( B + C )^n = 1/( 1 -B -C ) 
\ee

Repalcing $B$ by $w_B$, and $C$ by $w_C$, this immediately gives the
partition functions for $T > T_G$ as

\be
{\tilde {\cal Z}}(x,z,u~|~ .)= w_C /[x  ( 1 - w_B - w_C)]
\label{Ewbwc}
\ee
where the unspecified IS represented by the dot indicates a sum over all
possible IS.
To get ${\cal Z}(L,N)$, we have to evaluate the coefficient of $x^L z^N$
in the above expression. For large $L$ this varies as $\lambda^L$, where
$1/\lambda=x_c$, the singularity nearest to origin of (\ref{Ewbwc}). This
gives us $\lambda(z,u)$ as the solution of the quadratic equation 
\be
\lambda^2 -\lambda ( 1 + z u) + z( u-1) =0; 
\ee 
a result which is also obtainable directly by the transfer-matrix method.
The activity $z$ is determined using the condition $z \partial log \lambda
/ \partial z = N/L $.  For simplicity, we discuss below the special case
$N = L/2$. This corresponds to the activity being given by the equation
\be
z= 1/u, {\rm ~~for~~} N/L = 1/2. 
\ee 

In the language of spin models, this case corresponds to a simple Ising
chain with no external field. It is straightforward to determine different
averages, and correlation functions.  Differentiating $\lambda$ with
respect to $u$, we get the average energy $E$ as a function of $u$ 
\be 
E = - \frac{J \sqrt{u}}{2 (\sqrt{u} + 1)},{\rm ~for~} T > T_G. 
\ee

Similarly, we calculate the average number of substrings of the type $0
1^n 0$ per site. For $n =0$, this number is $1/(2 \lambda)$. for $n > 0$,
it is $1/(2 u \lambda^{n +1})$. Summing over the odd values of $n$, we get
that the fractional number of $n_B = N_B/L$ in a typical configuration is
given
by

\be
n_B = \frac{1}{2 ( 1 + 2 \sqrt{u})}, {\rm ~ for~~ } u < u_0.
\ee

Above $T_G$, the system can explore all configurations with different IS,
and the average value of $N_B$ decreases with decreasing temperature. As
we cool the system just below $T_G$, transitions that change IS are no
longer possible. Hence the sytem will remain in one sector, and the
average value of $n_B$ remains constant

\be
n_B  =  \frac{1}{2 ( 1 + 2 \sqrt{u_0})}, {\rm ~ for~ } u > u_0
\label{Enb}
\ee
where $u_0 = exp( J/k T_G)$.  There are $(L-N)!/( L - N -N_B)! N_B!$
equivalent sectors. The available phase space to the system shrinks by
this factor as the temperature falls below $T_G$.  The logarithm of this
quantity gives us the component of entropy that gets frozen at the glass
transition. In our special case $N = L/2$, this expression simplifies, and
the frozen entropy per site is

\be
\Delta S_{fr} = \frac{1}{2} H( 2 n_B)
\ee
where $H(x) = -x log x -( 1 - x) log (1- x)$.

For $T < T_G$, we have to calculate the pico-canonical partition function
in one of these sectors. This is easily evaluated using Eq.(\ref{Ezis}).
Since it only depends on $N_B$ and $N_C$, and these have only ${\cal
O}(\sqrt{L})$ fluctuations between different sectors, we see that the
pico-canonical free energy in each of these sector is the same. Averaging
over different sectors is then trivial, and has no effect. The average
energy is a continuous function of temperature at $T_G$.

Within a given sector, the relative weights of different configurations
are functions of temperature, and the average energy and entropy decrease
with temperature. The temperature dependence of average energy can be
determined from Eq.(\ref{Ezis}), or directly as follows:  There are
exactly $(L-N)$ zeroes in a configuration. Let the number of type $A$
particles that are followed by a type-$C$ particle be $K$. [We shall
include in this count a type $A$ particle at the end of chain.] In the
configuration, the energy of the configuration is $ ( -N +N_B + K) J$.
There are $(N_C+1)!/K! ( N_C -K+1)!$ ways of putting the $K$ particles
just left of $N_C +1$ symbols (treating the end mark as a $0$). Then, for
any such choice, we have to distribute $ (N_A -K)$ additional dimers left
of these $N_B + K$ symbols. This is the standard problem of distributing
$m$ identical balls in $n$ boxes, with more than one ball in a box
allowed. The answer is $(m+n-1)!/[m!(n-1)!]$. In our problem, the dimers
can be distributed in $ ( N_A + N_B -1)!/( N_A -K)!( N_B + K-1)$ ways.
This gives us

\be
{\cal Z}(L,N ~|~{\cal I})=\sum_{K=0}^{N_A} \frac { ( N_A+N_B-1)!
(N_C+1)!}{(
N_A -K)!( N_B+K-1)! K! ( N_C-K+1)!} u^{N-N_B-K}
\ee
As a simple check, we can verify that this correctly gives the partition
function of the sector shown in Fig.1.

In the limit of large $L$, we can use steepest descent to evaulate this
sum. The maximum contribution comes from $K = k L$, where $k$ satisfies
the equation

\be
k(n_B +k) = (1/ 2 u) ( 1 - n_B  - 2 k) ( 1/2 - n_B - k)
\ee
where $n_B$ is given Eq. (\ref{Enb}). For temperature tending to zero, $u$
tends to infinity, and $k$ tends to zero. As $k$ is equal to the excess
energy at temperature $T$ above the zero-temperature value, this gives us
the energy density at any temperature in the low-temperature phase.

  The minimum energy corresponds to $k=0$. We see that even at $T=0$, we
have a residual energy $ N_B J$ above the ground state energy. Also, in
addition to the frozen entropy, there is also a contribution to the zero
temperature entropy coming from the macroscopic degeneracy of the ground
state within the sector.

\section{Concluding Remarks}

In this paper, we described a simple model where the phase space shows
many-sector decomposition in the low temperature phase. We showed that
this sector decomposition can be characterized fully in terms of a
constant of motion called the irreducible string.  The exact partition
function can be calculated easily in each of these sectors. The relative
weights of different sectors in calculation of other observable averages
can also be calculated similarly within the model. In our model, it
depends only on the temperature at which ergodicity breaking occurs. More
generally, it would depend on the history of the sample.
 
In calculating thermodynamical quantities like the average energy, we
found that different sectors are not so different. In particular, the free
energy depended only on $N_B$ and $N_C$, and not on the arrangement of
characters of the irreducible string. And so, the average over sectors was
same as value in one sector. This property of self-averaging is expected
to work for other quantities as well, e.g. correlation functions.

Of course, the breaking of ergodicity is put into the model `by hand'.
However, it allows us to see explicitly, in a simple setting, concepts
like frozen entropy, and disallowing slow processes completely makes
possible discussion of glassy states as equilibrium states of matter. This
is admittedly an idealization, but useful, and perhaps no different from
other idealizations like the thermodynamic limit, ideal heat-baths (which
exchange energy with the system, but otherwise do not perturb it) etc.
well-known in equilibrium statistical mechanics. The main advantage is
that the arbitrariness in defining the notion of `nearby' states is
avoided. The naive implementation of the latter makes the ensemble depend
on the initial state, and on the maximum distance allowed.

Some generalizations of the model are straightforward, {\it e.g.}
inclusion of next-nearest neighbor interactions. One can also allow some
other relaxation processes, like $ 0010 \rightarrow 0100$. These reduce
the number of disjoint sectors, but the number still grows  exponentially
with the volume of the system.  Work on more realistic models, or higher
dimensional cases seems like a promising area for further study.


\begin{thebibliography}{99}
\bibitem{footnote} This has to be defined as a function of a small number
of observables, after integrating over all other degrees of fredom.
\end{thebibliography}
\end{document}